\documentclass[11pt,english,a4paper,floatfix]{article}
\usepackage{jcappub}
\usepackage{multirow}
\usepackage{xcolor}

\pdfoutput=1
\usepackage{bm}
\usepackage{amsfonts}

\newcommand{\CAMB}{\textsc{camb}}
\newcommand{\logg}{\log_{10}\left[G_\text{eff}\text{MeV}^2\right]}

\newcommand{\lcdm}{\Lambda\rm CDM}
\graphicspath{{/}}
\begin{document}


\title{Massive neutrino self-interactions and inflation}

\author[a,b]{Shouvik Roy Choudhury,}
\author[c]{Steen Hannestad,}
\author[c]{Thomas Tram}

\affiliation[a]{Department of Physics, Indian Institute of Technology Bombay, Main Gate Road, Powai, Mumbai 400076, India}
\affiliation[b]{Inter-University Centre for Astronomy \& Astrophysics (IUCAA), Ganeshkhind, Post Bag 4, Pune 411007, India}
\affiliation[c]{Department of Physics and Astronomy, Aarhus University,
 DK-8000 Aarhus C, Denmark}

\emailAdd{shouvikroychoudhury@gmail.com}
\emailAdd{sth@phys.au.dk}
\emailAdd{thomas.tram@phys.au.dk}

\abstract{Certain inflationary models like Natural inflation (NI) and Coleman-Weinberg inflation (CWI) are disfavoured by cosmological data in the standard $\Lambda\textrm{CDM}+r$ model (where $r$ is the scalar-to-tensor ratio), as these inflationary models predict the regions in the $n_s-r$ parameter space that are excluded by the cosmological data at more than 2$\sigma$ (here $n_s$ is the scalar spectral index). The same is true for single field inflationary models with an inflection point that can account for all or majority of dark matter in the form of PBHs (primordial black holes). Cosmological models incorporating strongly self-interacting neutrinos (with a heavy mediator) are, however, known to prefer lower $n_s$ values compared to the $\Lambda\rm CDM$ model. Considering such neutrino self-interactions can, thus, open up the parameter space to accommodate the above inflationary models. In this work, we implement the massive neutrino self-interactions with a heavy mediator in two different ways: flavour-universal (among all three neutrinos), and flavour-specific (involving only one neutrino species). We implement the new interaction in both scalar and tensor perturbation equations of neutrinos. Interestingly, we find that the current cosmological data can support the aforementioned inflationary models at 2$\sigma$ in the presence of such neutrino self-interactions.}

\maketitle


\section{Introduction}

Inflation as a theory has been highly successful in addressing various issues in the Big Bang Cosmology, namely the Horizon problem (why the universe appears homogeneous and isotropic), the Flatness problem (why the universe doesn't seem to have a curvature), and the Magnetic monopole problem (why we don't find any magnetic monopoles in the universe) (see \cite{Vazquez:2018qdg} for brief introductions to these problems). There are, however, plethora of inflationary models in the literature. Two important cosmological parameters pertaining to inflationary cosmology are the scalar spectral index ($n_s$) and the tensor-to-scalar ratio ($r$). Cosmic Microwave Background (CMB) and Large Scale Structure (LSS) probes can constrain the $n_s-r$ parameter region and may thereby rule out various inflationary models based on their expected $n_s$ and $r$ values. Quantum fluctuations in the inflationary field lead to the scalar and tensor perturbations. Specifically, the scalar perturbations lead to density fluctuations in the constituents of the universe (i.e. radiation, matter etc) which can be probed by the Cosmic Microwave Background (CMB) and Large Scale Structure (LSS) probes. On the other hand, while the tensor perturbations, i.e., the primordial gravitational waves can contribute to the all the CMB anisotopies \cite{Tristram:2020wbi}, the strongest constraints to the tensor-to-scalar ratio come from the CMB B-mode polarization \cite{Bicep:2021xfz}. 

The primordial scalar and tensor power spectra are usually parameterized as: $\mathcal{P}_s = A_s(k/k_*)^{n_s - 1}$ and $\mathcal{P}_{t} = A_t (k/k_*)^{n_t}$, respectively, with the tensor-to-scalar ratio $r \equiv A_t/A_s$. 
In the $\Lambda\textrm{CDM}+r_{0.05}$ model (where $r_{0.05}$ is the tensor-to-scalar ratio at the pivot scale of $k_*=0.05$ Mpc$^{-1}$), we obtained $n_s = 0.965 \pm 0.004$ (68\% C.L.) and $r_{0.05}< 0.034$ (95\% C.L.), using Planck 2018 temperature and polarisation data \cite{Planck:2018jri} combined with the latest publicly available CMB B mode data from BICEP/Keck collaboration \cite{Bicep:2021xfz}.  Unless otherwise specified, we shall use $r$ and $r_{0.05}$ interchangeably in this paper. 

A slow roll inflationary model can be described by a Lagrangian of the following form:
\begin{equation}
\mathcal{L} = \frac{1}{2} g^{\mu\nu}\partial_{\mu}\phi\partial_{\nu}\phi - V(\phi),
\end{equation}

where $\phi$ is the inflaton field and $V(\phi)$ describes its potential. The slow roll parameters are defined as:
\begin{equation}
\epsilon(\phi) \equiv \frac{m_{\rm pl}^2}{16\pi}\left(\frac{V^{'}}{V}\right)^2; ~~~~~ \eta(\phi) \equiv \frac{m_{\rm pl}^2}{8\pi}\left(\frac{V^{''}}{V}\right).
\end{equation}

Here $V^{'} \equiv \frac{dV}{d\phi}$, $V^{''} \equiv \frac{d^2V}{d\phi^2}$, and $m_{\rm pl}$ is the Planck mass. Our interest is in the phenomenological parameters that are cosmological observables, i.e. $n_s$ and $r$. These can be written in terms of the slow roll parameters as:

\begin{equation}
n_s = 1 - 6\epsilon (\phi_{s}) + 2\eta (\phi_{s}); ~~~~ r = 16\epsilon ({\phi_{s}}).
\end{equation}
Here $\phi_s$ denotes the value of the  $\phi$ field 40 to 60 e-folds before the end of inflation, as the fluctuations that are observable in the CMB are created during that time \cite{Baumann:2009ds}. Conventionally, we consider that inflation ends when the slow roll parameter $\epsilon({\phi_e}) = 1$, where $\phi_e$ denotes the value of $\phi$ at the end of inflation. The number of e-folds parameter is given by:

\begin{equation}\label{eq 1.4}
N_{*} \simeq -\frac{8\pi}{m_{\rm pl}^2}\int_{\phi_s}^{\phi_e} \frac{V}{V^{'}} d\phi.
\end{equation} 

Given a potential $V(\phi)$ and a particular choice of $N_{*}$, it is straightforward to calculate $\phi_e$, and then it is easy to calculate $\phi_s$ from equation \ref{eq 1.4}, and hence one can calculate the predicted $n_s$ and $r$ values.

In this work we are interested in two particular inflationary models: the Natural inflation (NI) \cite{Freese:1990rb,Adams:1992bn}, and the Coleman-Weinberg Inflation (CWI) \cite{Linde:1981mu,Albrecht:1982mp}. The potentials for these two models are given as follows:
\begin{eqnarray}
V_{\rm NI} (\phi) = \lambda^4 \left(1 + \mathrm{cos}\left(\frac{\phi}{g}\right)\right);\\
V_{\rm CWI}(\phi) = A \phi^4\left[\mathrm{ln}\left(\frac{\phi}{f}\right) - \frac{1}{4}\right]+\frac{Af^4}{4}.
\end{eqnarray}

Here $\lambda$, $g$, $A$, and $f$ are parameters in the models. It can be shown that for the CWI inflation, in the small field inflation regime, i.e. $(\phi/f)\ll 1$, we have $N_{*}\simeq 3/(1-n_s)$, and $r\simeq 0$ \cite{Barenboim:2019tux}.

The latest Planck results rule out NI at more than 2$\sigma$ \cite{Planck:2018jri} in the $\Lambda\textrm{CDM}+r$ model, with Planck 2018 CMB anisotropies \cite{Aghanim:2018eyx} combined with the older BICEP/Keck CMB B mode data, BK15 \cite{BICEP2:2018kqh}. The CWI model has been ruled out at more than 2$\sigma$ with Planck data, much before than NI \cite{Barenboim:2013wra}. In our work, we have also found that both the models are ruled out at more than 2$\sigma$ with latest cosmological datasets in the $\Lambda\textrm{CDM}+r$ model.

We are also interested in single field inflationary models with an inflection point that can produce all or majority of the dark matter content in the universe in the form of primordial black holes (PBHs) \cite{Ballesteros:2017fsr,Ballesteros:2020qam,Mishra:2019pzq}. Such models require a spectral index value of $n_s \simeq 0.95$ \cite{Ballesteros:2020qam} which is much lower than the $\Lambda\textrm{CDM}$ bounds and hence disfavoured at more than 2$\sigma$ as well. Hereafter, we shall refer these models as PBH DM related inflationary models.

Augmenting the cosmological model with non-standard self-interactions among all 3 neutrinos with a heavy mediator has been shown to bring back NI and CWI within the 1$\sigma$ region in the $n_s-r$ plane \cite{Barenboim:2013wra}, using Planck 2015 CMB temperature anisotropies, low-multipole polarization, and lensing \cite{Ade:2015xua}. It is important to note that both NI and CWI models can be reconciled with cosmological data if the chosen cosmological model prefers lower values of $n_s \simeq 0.93-0.94$ instead of the preferred region by the $\lcdm$ model, which is around $n_s\simeq0.965$. At the same time, the PBH DM related inflationary models can be accommodated for $n_s \simeq 0.95$. Preference for such lower $n_s$ values is exactly what is possible with strongly self-interacting neutrinos, where the coupling strength is $\sim$$10^9$ times that of the weak interaction. In this work we revisit the self-interacting neutrino model in the context of these aforesaid inflationary models and test this model with new datasets. Below we briefly introduce the massive neutrinos and the interaction model.      

Neutrinos are massless in the standard model of particle physics, but terrestrial neutrino oscillation experiments~\cite{Fukuda:1998mi,Ahmad:2002jz} have confirmed that there are 3 non-degenerate neutrino mass eigenstates (with at least two of the masses being small but non-zero). These mass eigenstates are quantum super-positions of their flavour eigenstates. Cosmological data is sensitive to the neutrino energy density, which is proportional to the sum of neutrino masses, $\sum m_{\nu}$ when all the neutrinos become non-relativistic. At present, the bound on $\sum m_{\nu}$ is around $\sum m_{\nu} \lesssim 0.12$ eV (95\% C.L.)~\cite{Aghanim:2018eyx,RoyChoudhury:2019hls,Choudhury:2018byy,Alam:2020sor,Vagnozzi:2017ovm,Ivanov:2019hqk,Tanseri:2022zfe}, while the most stringent bounds quoted in literature is $\sum m_{\nu} < 0.09$ eV (95\% C.L.) \cite{DiValentino:2021hoh,diValentino:2022njd,Palanque-Delabrouille:2019iyz} under the assumption of a $\Lambda\textrm{CDM}+\sum m_{\nu}$ cosmology with 3 degenerate neutrino masses. This bound can relax up to a factor of 2 or more in extended cosmologies \cite{RoyChoudhury:2019hls,Choudhury:2018byy}. However, physically motivated restrictions to the parameter space can lead to stronger bounds than the $\Lambda\rm CDM$ cosmology \cite{RoyChoudhury:2018vnm,Vagnozzi:2018jhn}. Impact of neutrino properties like mass and energy density on the $n_s-r$ plane has been discussed in \cite{Gerbino:2016sgw}. See \cite{MoradinezhadDizgah:2021upg,Brinckmann:2018owf} for forecasts on constraints on neutrino masses from future cosmological data.

There are a plethora of models that have been proposed to explain the generation of neutrino masses. Here we consider the majoron model where we consider the neutrinos to be Majorana particles, and the $U(1)_{B-L}$~\cite{Gelmini:1980re,Choi:1991aa,Acker:1992eh,Chikashige:1980ui,Georgi:1981pg} symmetry is spontaneously broken, leading to a new Goldstone boson, the majoron. We denote the majoron by $\Phi$. It couples to the neutrinos via the Yukawa interaction~\cite{Oldengott:2017fhy,Oldengott:2014qra},

\begin{equation}
\mathcal{L}_{\rm int} = g_{ij} \bar{\nu_i} \nu_j \Phi + h_{ij}  \bar{\nu_i} \gamma_5 \nu_j \Phi,
\end{equation} \label{eq 1.7}
where $\nu_i$ is a left-handed neutrino Majorana spinor, $g_{ij}$ and $h_{ij}$ are the scalar and pseudo-scalar coupling matrices, respectively. The indices $i,j$ are used to label the neutrino mass eigenstates. We note here that in general this kind of interaction is not limited to the majoron-like model of neutrino mass generation. For instance, $\phi$ can be linked to the dark sector~\cite{Barenboim:2019tux}.

In this paper we consider the two scenarios: i) a flavour universal scenario (all 3 neutrinos interacting), ii) a flavour specific scenario (only 1 neutrino species interacting). In the flavour universal scenario we take $g_{ij} = g \delta_{ij}$ and $h_{ij} = 0$, where $\delta_{ij}$ is the Kronecker delta. Thus, in both flavour and mass basis $g_{ij}$ has the same form. Such a flavour universal interaction scenario may not be realistic for particle physics models, but it provides a simple method of testing the sensitivity of cosmological data to such neutrino-majoron interactions. At the same time, we note that the flavour universal interaction scenario is strongly constrained by particle physics experiments, and self-interactions among only the $\tau$ neutrinos is the least constrained \cite{Blinov:2019gcj,Brdar:2020nbj,Lyu:2020lps,Berbig:2020wve}. This motivates us to consider the second scenario which is flavour specific where we consider only one neutrino species interacting. Here we consider $g_{ij}$ to be diagonal with only one non-zero component, i.e., $g_{ij} = g\delta_{kk}\delta_{ij}$, where $k$ is either 1, 2, or 3 (no sum over $k$ is implied). We note here that unlike the flavour universal case, here a diagonal $g_{ij}$ in the flavour basis with only one non-zero component $g_{\tau\tau}$ (since only $\tau$ neutrinos are interacting among each other) shall not translate to a diagonal $g_{ij}$ in the mass basis with only one non-zero component. However, we expect the non-diagonal terms or other diagonal terms in the mass-basis $g_{ij}$ to be small considering that we are dealing with small neutrino masses, as the neutrino mass bounds from cosmological data are quite stringent as mentioned above, and these mass bounds almost remain unchanged even with the presence of strong neutrino self-interactions \cite{RoyChoudhury:2020dmd}. Thus we expect that the approximation of only one mass eigenstate self-interaction to represent the self-interaction among the $\tau$ neutrinos to be a good approximation.

We choose the mass of the scalar $m_{\Phi}$ to be  much larger than the energies of neutrinos during the CMB epoch, so as to be able to consider the interaction to be, effectively, a 4-fermion interaction during and after the CMB epoch, and the $\Phi$ particles would have decayed away. A mass of  $m_{\Phi} > 1$ keV should be enough to ensure this \cite{Blinov:2019gcj}, however one might consider $m_{\Phi} > 1$ MeV to avoid constraints from the Big Bang Nucleosynthesis as well. We emphasize here that such a scenario is not limited to scalar particles, and in fact all the results and conclusions in this paper will be applicable for a heavy vector boson as well \cite{Ohlsson:2012kf,Archidiacono:2013dua}. 

Now we can treat the interaction Lagrangian in equation \ref{eq 1.7} as a $\nu \nu \rightarrow \nu \nu$ self-interaction with a self-interaction rate per particle $\Gamma \sim g^4 T_{\nu}^5/m_{\phi}^4 = G_{\rm eff}^2 T_{\nu}^5$, where $G_{\rm eff} = g^2/m_{\phi}^2$ is the effective self-coupling~\cite{Oldengott:2017fhy}. In such a scenario, the neutrinos as usual decouple from the primordial plasma at the decoupling temperature $T \sim 1$ MeV. This happens when the weak interaction rate falls below the Hubble rate, i.e., $\Gamma_{\rm W} < H$, with $\Gamma_{\rm W} \sim G_{\rm W}^2 T_{\nu}^5$. Here $G_{\rm W} \simeq 1.166 \times 10^{-11} \textrm{MeV}^{-2}$ is the standard Fermi constant. However, after decoupling from the primordial plasma, the neutrinos continue to scatter among themselves, assuming $G_{\rm eff} > G_{\rm W}$. They continue to do so until the self-interaction rate $\Gamma$ falls below the Hubble rate, and after that they will free-stream. So by increasing  $G_{\rm eff}$, one can further delay the neutrino free-streaming. Very strong interactions like $G_{\rm eff} \simeq 10^9 G_{\rm W}$ can delay free-streaming till matter radiation equality. 
\footnote{When $m_\Phi \sim T$ or smaller the phenomenology of the model changes significantly: The system undergoes recoupling instead of decoupling, and a new population of $\Phi$ particles can be built up from neutrino pair annihilation. We refer the reader to e.g.\ Refs.\ \cite{Archidiacono:2013dua,Forastieri:2015paa,Forastieri:2019cuf,Archidiacono:2020yey,Escudero:2021rfi,Escudero:2019gvw,Corona:2021qxl,Venzor:2022hql} for a more detailed discussion. See also \cite{Chacko:2020hmh,Taule:2022jrz,Blinov:2020hmc,Escudero:2019gfk,Park:2019ibn,Chang:2022aas,Das:2022xsz,Esteban:2021tub,Huang:2021dba,Mazumdar:2019tbm,Sung:2021swd,Esteban:2021ozz,Kharlanov:2020cti,Venzor:2020ova} for discussions in the related fields.}
   
See~\cite{RoyChoudhury:2020dmd,Oldengott:2017fhy,Barenboim:2019tux,Archidiacono:2013dua,Cyr-Racine:2013jua,Lancaster:2017ksf,Kreisch:2019yzn,Brinckmann:2020bcn,Das:2020xke,Mazumdar:2020ibx,Kreisch:2022zxp} for previous studies on cosmological constraints on $G_{\rm eff}$ (specifically, the $\logg$ parameter). Strong interactions like $G_{\rm eff} \simeq 10^9 G_{\rm W}$ are allowed in the CMB data mainly through a degeneracy present among $G_{\rm eff}$, the angular size of the sound horizon at the last scattering $\theta_s$, and the scalar spectral index $n_s$. This degeneracy leads to bimodal posterior distributions with distinct modes in these three parameters as well. Strong interactions due to a large $G_{\rm eff}$ pertain to a lack of anisotropic stress in the neutrino sector, the effect of which on the CMB power spectra can be compensated partially by increasing $\theta_s$. At the same time, increasing $G_{\rm eff}$ causes a gradual increase in the power in small scales of the CMB power spectrum which can be partially compensated by a smaller $n_s$~\cite{Oldengott:2017fhy}. 

As mentioned before, in the context of Natural Inflation, Coleman-Weinberg Inflation, and PBH DM related inflationary models, a smaller $n_s$ is quite useful. To put constraints on the $n_s$ - $r_{0.05}$ plane, one needs to introduce the tensor perturbation equations as well. In this work, we introduce modifications to both the scalar and tensor perturbation equations of neutrinos to take care of the effects of the self-interaction, in the CAMB code \cite{Lewis:2002ah}. The background equations remain unchanged as the $\Phi$ particles have decayed away for our epochs of concern and any possible changes in the neutrino temperature due to the decay is absorbed into the $N_{\rm eff}$ parameter. We work in the extended $\Lambda\textrm{CDM} + r_{0.05} + \logg + N_{\rm eff}$ + $\sum m_{\nu}$ model, where, for our purposes, $N_{\rm eff}$ is the effective number of neutrino species (in general, it constitutes any relativistic species other than photons, in the early universe). As noted before, we consider two scenarios: i) all 3 neutrino species self-interacting, and ii) only 1 neutrino species interacting. We test this model against the full Planck 2018 temperature and polarization likelihoods \cite{Aghanim:2019ame}, the latest CMB B mode data from BICEP/Keck collaboration \cite{Bicep:2021xfz} and with additional data from Planck 2018 lensing \cite{Aghanim:2018oex}, BAO and RSD measurements \cite{Alam:2016hwk,Ross:2014qpa,Beutler:2011hx}, and uncalibrated Type Ia Supernovae luminosity distance measurements \cite{Scolnic:2017caz}. We find that both the inflation models can be accommodated within 2$\sigma$ in the $n_s-r_{0.05}$ plane. Our results are different than the previous work in this area \cite{Barenboim:2019tux}, where the authors could reconcile the NI and CWI models with older data within 1$\sigma$. Our results thus add to the literature in a meaningful way and tighten the constraints on these two inflationary models in the presence of self-interacting neutrinos. We also find that the PBH DM related inflationary models can be accommodated at 2$\sigma$ as well.

The rest of the paper is structured as follows. In section~\ref{section:2} we present the modifications to the neutrino Boltzmann equations, the cosmological model parametrisation and priors and the analysis method adopted, as well as the cosmological datasets used in this paper. In section~\ref{section:3} we present the results of the analyses and in section~\ref{section:4} we conclude.

\section{Methodology}
\label{section:2}

As explained in the previous section, our work uses the neutrino self-interactions mediated by a heavy scalar. While this would mean that we shall use some specific coefficients in the interaction terms in the collisional Boltzmann equations for the neutrinos, our results will be generally applicable to other neutrino non-standard interactions with heavy mediators, e.g. a gauge boson. 

We implement the modified cosmological perturbation equations in the \CAMB{} code  \cite{Lewis:1999bs}. We consider the background equations to remain unchanged due to neutrino self-interactions, which is a superb approximation considering the heavy mediator decays away way before photon decoupling. For the flavour-independent case (hereafter ``3$\nu$-interacting'' case), the modifications to the perturbation equations apply to all the three neutrino species, while in the flavour-specific case (hereafter ``1$\nu$-interacting'' case), the modifications apply only to one of the three species. 

\subsection{Cosmological perturbation equations} 

To incorporate the self-interaction in the neutrino perturbation equations in \CAMB{}, we use the relaxation time approximation (RTA) that was first introduced in this context in~\cite{Hannestad:2000gt} (and first used for a treatment of self-interactions in light neutrinos in~\cite{Hannestad:2004qu}). In~\cite{Oldengott:2017fhy}, RTA was found to be very accurate and consistent when compared to the exact collisional Boltzmann equations. We emphasize here that we have implemented the modifications to both scalar and tensor perturbation equations.

\label{key}
In the scalar perturbation equations, these scattering interactions cause a damping in the Boltzmann hierarchy for multipoles $\ell\geq 2$. Following the notation in \cite{Ma:1995ey}, in the synchronous gauge, the collisional Boltzmann hierarchy for massive neutrino scalar perturbations is given by, 
\begin{eqnarray}
\label{eq:boltzman}
\dot{\Psi}_0 &=& -{qk\over \epsilon}\Psi_1
+{1\over 6}\dot{h} {d\ln f_0\over d\ln q}
\,, \nonumber\\
\dot{\Psi}_1 &=& {qk\over 3\epsilon} \left(\Psi_0
- 2 \Psi_2 \right) \,, \nonumber\\
\dot{\Psi}_2 &=& {qk\over 5\epsilon} \left(
2\Psi_1 - 3\Psi_3 \right)
- \left( {1\over15}\dot{h} + {2\over5} \dot{\eta} \right)
{d\ln f_0\over d\ln q} + \alpha_2 \dot{\tau}_\nu \Psi_2\,,\\
\dot{\Psi}_l &=& {qk \over (2l+1)\epsilon} \left[ l\Psi_{l-1}
- (l+1)\Psi_{l+1} \right] + \alpha_\ell \dot{\tau}_\nu \Psi_l \,,
\quad l \geq 3 \,. \nonumber
\end{eqnarray}

where $\alpha_\ell \dot{\tau}_\nu \Psi_l$ are the damping terms for $l\geq 2$. Needless to say, these terms do not appear for the non-interacting neutrinos in the flavour-specific interaction case. Here $\dot \tau_\nu \equiv -a G_{\rm eff}^2 T_\nu^5$ is the opacity for the neutrino self-interactions with a heavy mediator, and $\alpha_l$ ($l>1$) are coefficients of order unity that depend on the interaction model. We use $\alpha_l$ values from equation\ 2.9 in~\cite{Oldengott:2017fhy} for the scalar mediator, i.e., we use $\alpha_2 = 0.40$, $\alpha_3 = 0.43$, $\alpha_4 = 0.46$,  $\alpha_5 = 0.47$, $\alpha_{l \geq 6} = 0.48$. For neutrino tensor perturbation equations we follow a similar procedure and add similar damping terms to the equations in the CAMB code \cite{Lewis:1999bs}. However, we set all $\alpha_l = 1$ ($l>1$), instead of choosing model specific values, since these model dependent coefficients for tensor perturbation equations require a separate calculation. We have verified that the CMB B-mode spectrum due to primordial tensor perturbations goes through only a minor change when we vary $\alpha_l$ from 0.4 to 1. Thus, setting all $\alpha_l = 1$ in the neutrino tensor perturbation equations is only likely to produce some minor shifts in the value of $\logg$ and hence, is not of major concern.

In the very early universe, a tight coupling approximation (TCA) was employed in our codes, where only the two lowest moments are non-zero. This was done since the collisional Boltzmann equations for neutrinos are not easy to solve in the very early universe. This approximation is switched off early enough (when $|\dot \tau_\nu|/\mathcal{H} < 1000$, where $\mathcal{H}$ is the conformal Hubble parameter), so that it does not bias our results. 

\subsection{Cosmological model: parametrization and priors}

Our cosmological model of interest consists of an extended $\Lambda$CDM model that includes the tensor-to-scalar ratio $r_{0.05}$, and massive neutrinos' mass sum, energy density, and interaction strength parametrized with $\sum m_{\nu}$, $N_{\rm eff}$, and  $G_{\rm eff}$ respectively. Note that the model with all 3 interacting neutrinos (3$\nu$-interacting, flavour independent) and the model with only 1 interacting neutrino (1$\nu$-interacting, flavour specific) both have the same parameters.

Thus, both our 3$\nu$-interacting and 1$\nu$-interacting cosmological models can be represented by the same following parameter vector: 
\begin{equation}\label{eq:model}
{\bm \theta} = \{\Omega_{\rm c} h^2,\Omega_{\rm b} h^2,100\theta_{MC},\tau,
\ln(10^{10}A_{s}),n_{s}, r_{0.05}, \sum m_\nu, N_{\rm eff}, \textrm{log}_{10} \left[G_{\rm eff} \textrm{MeV}^2\right]\}.
\end{equation}
Here, the first six parameters pertain to the $\Lambda$CDM model. $\Omega_{\rm c} h^2$ and $\Omega_{\rm b} h^2$ are the physical densities at present ($z=0$) for cold dark matter and baryons respectively, $100\theta_{MC}$ is the parameter used by CosmoMC as an approximation for the angular size of the sound horizon, $\theta_s$. We have $\tau$ as the reionization optical depth, and $\ln(10^{10}A_s)$ and $n_s$ are, respectively, the amplitude and spectral index of the primordial scalar fluctuations, at a pivot scale of $k=0.05 \rm h~Mpc^{-1}$. 
The tensor-to-scalar ratio $r_{0.05}$ is an important parameter for inflationary models, and we also consider a pivot scale of $k=0.05 \rm h~Mpc^{-1}$ for this parameter. 
 
We assume a degenerate hierarchy of neutrino masses, i.e., each neutrino has a mass of $m_\nu = \frac{1}{3}\sum m_\nu$. Currently there is no conclusive evidence for preference of normal or inverted hierarchy of neutrino masses \cite{RoyChoudhury:2019hls,Gariazzo:2022ahe,Gariazzo:2018pei,Heavens:2018adv,Lattanzi:2017ubx,Gerbino:2016ehw}, and thus the degenerate approximation is okay to be used as far as current or even future cosmological data is concerned \cite{Archidiacono:2020dvx,Mahony:2019fyb}. We use a flat prior on $\textrm{log}_{10} \left[G_{\rm eff} \textrm{MeV}^2\right]$ instead of $G_{\rm eff}$ as it allows us to vary the parameter over multiple orders of magnitude. Inside the logarithm $G_{\rm eff}$ is expressed in units of MeV$^{-2}$.

Note that in this work, we divide the $N_{\rm eff}$ equally among the 3 neutrinos. So in the 3$\nu$-interacting model, all of the $N_{\rm eff}$ is associated with the self-interacting neutrinos, whereas in the 1$\nu$-interacting model, only $N_{\rm eff}/3$ is associated with self-interacting neutrinos and the rest corresponds to free-streaming neutrinos.

The priors on each model parameter is listed in table~\ref{tab:priors}. As the posterior for $\textrm{log}_{10} \left[G_{\rm eff} \textrm{MeV}^2\right]$ is bimodal for the full range $[-5.5, -0.1]$ \cite{RoyChoudhury:2020dmd}, to obtain parameter constraints pertaining to each individual mode, we split the prior range in two: $-5.5\rightarrow -2.3$ for the Moderately Interacting mode (denoted MI$\nu$) and $-2.3\rightarrow -0.1$ for the Strongly Interacting mode (denoted SI$\nu$). This is done for both the 3$\nu$-interacting and 1$\nu$-interacting cases.

We also perform analysis in the $\Lambda\textrm{CDM} + N_{\rm eff}+ \sum m_{\nu}+r_{0.05}$ model, as we want to compare the interacting models with the non-interacting case. We denote this model by NI$\nu$.

\begin{table}[t]
\caption{Uniform priors for all the cosmological model parameters.}
\label{tab:priors}
\begin{center}
\begin{tabular}{lr@{$\,\to\,$}l}
\hline
 Parameter & \multicolumn{2}{c}{Prior}\\
\hline
$\Omega_{\rm b}h^2$ & $0.019$ & $0.025$\\
$\Omega_{\rm c}h^2$ & $0.095$ & $0.145$\\
$100\theta_{MC}$ & $1.03$ & $1.05$\\
$\tau$ & $0.01$ & $0.1$\\
$n_s$ & $0.885$ & $1.04$\\
$\ln{(10^{10} A_s)}$ & $2.5$ & $3.7$\\
$r_{0.05}$ & $0$ & $0.3$ \\
$\sum m_\nu$ [eV] &  $0.005$ & $1$\\
$N_{\rm eff}$ & $2$ & $5$\\
$\textrm{log}_{10} \left[G_{\rm eff} \textrm{MeV}^2\right]$  & $-5.5$ & $-0.1$\\
\hline
\end{tabular}
\end{center}
\end{table}

\subsection{Datasets}\label{sec:datasets}

We use the full CMB temperature and polarisation data (i.e. TT, TE, EE + lowE) from the Planck 2018 public data release~\cite{Aghanim:2018eyx}. We simply denote this as Planck18. Specifically, TT denotes the low-$l$ and high-$l$ temperature power spectra, whereas TE denotes the high-$l$ temperature and E-mode polarisation cross-spectra, EE denotes the high-$l$ E-mode polarisation spectra, and lowE denotes the low-$l$ E mode polarisation spectra. Here we mention that we use the full Planck likelihood where all the nuisance parameters are varied along with the main model parameters. We also use the B-mode CMB power spectra data from the BICEP2/Keck array public data release \cite{Bicep:2021xfz} that includes observations up to 2018, and denote this simply as BK18. We always use Planck18 and BK18 together, and name this combination CMB. \\

~~~~~~~~~~~~~~~~~~~~~~~~~~~	  \textbf{CMB} $\equiv$ \textbf{Planck18} $+$ \textbf{BK18} \\

In addition to the CMB power spectra, we use an additional dataset combination which consists of Planck 2018 CMB lensing~\cite{Aghanim:2018oex}, BAO and RSD measurements from SDSS-III BOSS DR12~\cite{Alam:2016hwk}, additional BAO measurements from MGS \cite{Ross:2014qpa} and 6dFGS \cite{Beutler:2011hx}, and SNe~Ia luminosity distance measurements from the Pantheon sample~\cite{Scolnic:2017caz}. We denote this combination as EXT. \\

~~~~~~~~~~~~ \textbf{EXT} $\equiv$ \textbf{Planck 2018 lensing} $+$ \textbf{BAO} $+$ \textbf{RSD} $+$ \textbf{SNe~Ia} \\

\subsection{Parameter sampling and analysis}

To effectively sample the bimodal posterior distribution and to calculate Bayesian evidences, we use the nested sampling package Polychord~\cite{Handley:2015vkr,Handley:2015fda} added to CosmoMC~\cite{Lewis:2002ah,Lewis:2013hha}. This extension to CosmoMC is known as CosmoChord \cite{Handley}. We used high settings of 4000 live points with boost\_posterior = 3 for the runs that incorporated the full range of $\textrm{log}_{10} \left[G_{\rm eff} \textrm{MeV}^2\right]$, i.e., $-5.5$ $\rightarrow$ $-0.1$. This is to compute accurate posterior distributions and Bayesian evidences for the bimodal posterior scenario. The posterior distributions from the non-interacting case (NI$\nu$), and the moderately interacting (MI$\nu$), and the strongly interacting case SI$\nu$ are all unimodal, and thus require a less intensive settings of 2000 live points and boost\_posterior = 0. We use HMcode \cite{Mead:2020vgs} (included with the CosmoChord package) to handle non-linearities. We use GetDist~\cite{Lewis:2019xzd} to generate the parameter bounds and posterior plots.

\section{Results}
\label{section:3}

\begin{figure}[tbp]
    \includegraphics[width=0.45\textwidth]{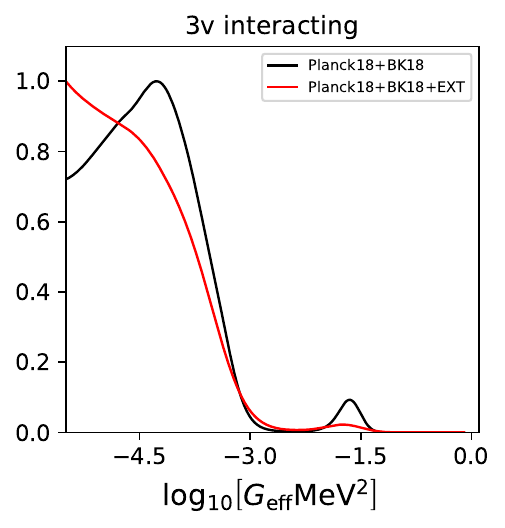}
    \hfill
    \includegraphics[width=0.45\textwidth]{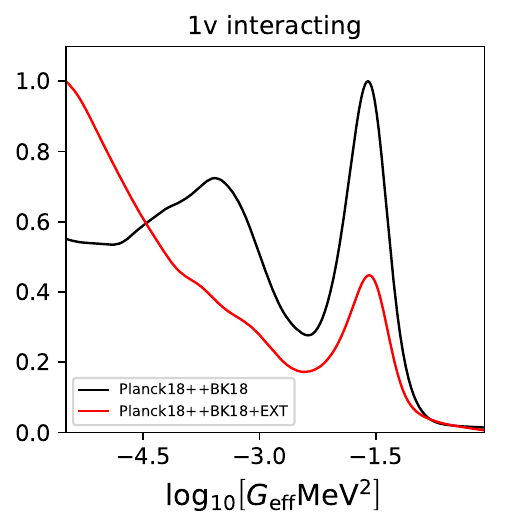}
    \caption{\label{fig:1} Posterior distributions of $\logg$ from the runs with the full range of coupling strengths $\logg$, for the two cases: all 3 neutrinos interacting (3$\nu$-interacting), and only 1 neutrino interacting (1$\nu$-interacting). We have provided the plots for two dataset combinations: Planck18+BK18 and Planck18+BK18+EXT. Compared to the 3$\nu$-interacting case, the 1$\nu$-interacting case has a much more pronounced SI$\nu$ peak. Details about models and datasets are given in section \ref{section:2}. }
\end{figure}

\begin{figure}[tbp]
	\includegraphics[width=0.45\textwidth]{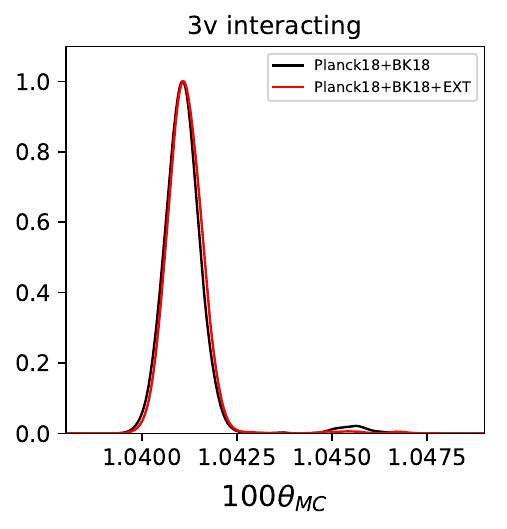}
	\hfill
	\includegraphics[width=0.45\textwidth]{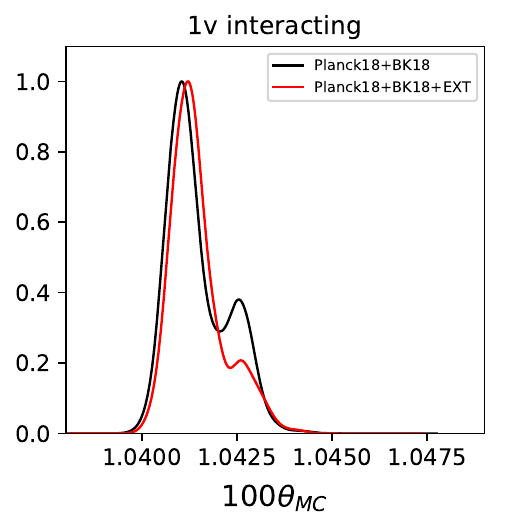}
	\caption{\label{fig:2} Posterior distributions of $100\theta_{MC}$ from the runs with full range of the coupling strength $\logg$, for the two cases: all 3 neutrinos interacting (3$\nu$-interacting), and only 1 neutrino interacting (1$\nu$-interacting). We have provided the results for two dataset combinations: Planck18+BK18 and Planck18+BK18+EXT. Compared to the 3$\nu$-interacting case, the 1$\nu$-interacting case has a much smaller shift in the SI$\nu$ peak, i.e., in the 1$\nu$-interacting case, the two peaks overlap. Details about models and datasets are given in section \ref{section:2}.}
\end{figure}

Our main results from the cosmological parameter estimation runs are tabulated in table~\ref{tab:evidence}, and visualised in figures~\ref{fig:1}--\ref{fig:6}.
Below we briefly summarize our results regarding the cosmological parameters:

\begin{table}
	\renewcommand*{\arraystretch}{1.2}

\small
\begin{tabular} { |l r|  c c | c c|}

\hline
 &  & \multicolumn{2}{c|}{ 3$\nu$ interacting}  & \multicolumn{2}{c|}{1$\nu$ interacting}\\

& & CMB &  CMB+EXT &  CMB  &  CMB+EXT \\
\hline
\multirow{3}{*}{{\boldmath$n_s            $} }&NI$\nu$& $0.959\pm0.009$ & $0.965\pm 0.006$  & $0.959\pm0.009$ & $0.965\pm 0.006$ \\
&MI$\nu$& $0.960^{+0.008}_{-0.009}$ & $0.963^{+0.008}_{-0.007}   $ & $0.959\pm0.009   $ & $0.964^{+0.007}_{-0.006}   $\\
&SI$\nu$& $0.930 \pm 0.008$ & $0.930\pm0.006$ & $0.950^{+0.008}_{-0.009}$ & $0.954\pm0.007   $\\
\hline

\multirow{3}{*}{{\boldmath$r_{0.05}$} }&NI$\nu$& $<0.034$ & $<0.037$ & $<0.034$ & $<0.037$ \\
&MI$\nu$& $<0.033$ & $<0.034$ & $<0.034$ &$<0.037$\\
&SI$\nu$& $<0.038$ & $<0.037$ & $<0.034$ & $<0.035$\\

\hline
\multirow{3}{*}{{\boldmath$\log_{10} \left[G_\mathrm{eff}\mathrm{MeV}^2\right]$} }&NI$\nu$&        $-$                      &         $-$                     &  $-$                            &    $-$                         \\
&MI$\nu$& $<-3.52                   $ & $<-3.32                   $ & $<-2.99                  $ & $<-2.4                   $\\
&SI$\nu$& $-1.67^{+0.16}_{-0.12}     $ & $-1.70^{+0.17}_{-0.10}     $ & $-1.67^{+0.38}_{-0.33}     $ & $-1.59^{+0.33}_{-0.39}     $\\

\hline
\multirow{3}{*}{{\boldmath$100\theta_{MC}$} }&NI$\nu$&        $1.04112\pm 0.00045$                      &         $1.04112\pm 0.00041$                    &  $1.04112\pm 0.00045$                           &    $1.04112\pm 0.00041$                         \\
&MI$\nu$& $1.04106\pm0.00045$ & $1.04111\pm0.00044$ & $1.04109\pm0.00047$ & $1.04109\pm0.00042$\\
&SI$\nu$& $1.04564^{+0.00077}_{-0.00053}      $ & $1.04554^{+0.00083}_{-0.00051}     $ & $1.04243^{+0.00061}_{-0.00054}     $ & $1.04262^{+0.00059}_{-0.00054}     $\\

\hline

\multirow{3}{*}{{\boldmath$\Omega_b h^2$} }&NI$\nu$& $0.02224\pm0.00023$ & $0.02238\pm0.00017$  & $0.02224\pm0.00023$ & $0.02238\pm0.00017$\\
&MI$\nu$& $0.02226^{+0.00023}_{-0.00021}$ & $0.02233\pm0.00019$ & $0.02226\pm0.00024$ & $0.02237^{+0.00017}_{-0.00018}$\\
&SI$\nu$& $0.02235\pm0.00023$ & $0.02236\pm0.00018$ & $0.02226^{+0.00022}_{-0.00024}$ & $0.02236^{+0.00018}_{-0.00020}$\\

\hline
\multirow{3}{*}{{\boldmath$\Omega_c h^2$} }&NI$\nu$&$0.1184^{+0.0030}_{-0.0033}$  & $0.1185\pm0.0027$ & $0.1184^{+0.0030}_{-0.0033}$ & $0.1185\pm0.0027$\\
&MI$\nu$& $0.1191^{+0.0028}_{-0.0029}$ & $0.1184^{+0.0028}_{-0.0030}$ & $0.1186\pm0.0032$ & $0.1189\pm0.0028$\\
&SI$\nu$& $0.1167^{+0.0030}_{-0.0032}$ & $0.1160\pm0.0027$ & $0.1186\pm0.0031$ & $0.1185^{+0.0030}_{-0.0035}$\\

\hline
\multirow{3}{*}{{\boldmath$\Sigma m_\nu$ [eV]} }&NI$\nu$& $< 0.235                  $ & $<0.119$ & $< 0.235                   $ & $<0.119$ \\
&MI$\nu$& $< 0.248                  $ & $< 0.121                   $ & $< 0.252                 $ & $< 0.120                  $\\
&SI$\nu$& $< 0.276                   $ & $< 0.161                   $ & $< 0.268                   $ & $< 0.145                   $\\

\hline
\multirow{3}{*}{{\boldmath$N_\mathrm{eff}$} }&NI$\nu$& $2.91\pm0.20      $ & $2.99\pm0.16$ & $2.91\pm0.20$ & $2.99\pm0.16$\\
&MI$\nu$& $2.96\pm0.19      $ & $2.97^{+0.17}_{-0.19}      $ & $2.93\pm0.21      $ & $3.01\pm0.16      $\\
&SI$\nu$& $2.78^{+0.18}_{-0.20}      $ & $2.76^{+0.15}_{-0.17}      $ & $2.91^{+0.19}_{-0.21}      $ & $2.96^{+0.18}_{-0.21}      $\\

\hline
\multirow{3}{*}{$H_0$ [km/s/Mpc] }&NI$\nu$& $66.2^{+1.7}_{-1.6}        $ & $67.4\pm 1.0$ & $66.2^{+1.7}_{-1.6}$ & $67.4\pm1.0$\\
&MI$\nu$& $66.4^{+1.8}_{-1.5}        $ & $67.2\pm1.1        $ & $66.3^{+1.9}_{-1.7}        $ & $67.5\pm1.0 $\\
&SI$\nu$& $66.7^{+1.8}_{-1.7}             $ & $66.9^{+1.0}_{-1.1}        $ & $66.4^{+1.8}_{-1.7}        $ & $67.5^{+1.1}_{-1.3}       $\\
\hline
\multirow{3}{*}{$Z/Z_{\text{NI}\nu}$} & NI$\nu$ & $1.000$ & $1.000$ & $1.000$ & $1.000$ \\
&MI$\nu$& 1.547 & 0.342 & 1.785 & 1.318\\
&SI$\nu$& 0.079 & 0.102 & 1.279 & 1.158\\
\hline
\multirow{3}{*}{$-2 \left[\log \left( \mathcal{L} / \mathcal{L}_{\text{NI}\nu} \right) \right]$ \kern-1em} & NI$\nu$ & $0$ & $0$ & $0$ & $0$ \\
&MI$\nu$& -1.58 & 1.02 & -1.89 & 1.40 \\
&SI$\nu$& 5.59 & 3.80 & -1.52 &  -1.19 \\
\hline
\end{tabular}

	\caption{\label{tab:evidence} Parameter constraints in the non-interacting model (NI$\nu$), moderately interacting model (MI$\nu$) and the strongly interacting model (SI$\nu$). For MI$\nu$ and SI$\nu$, we have the two scenarios: i) all 3 neutrinos interacting, and ii) only 1 neutrino interacting. The constraints are reported for two different dataset combinations: CMB and CMB+EXT. Marginalized constraints are given at 1$\sigma$, whereas upper bounds are given at 2$\sigma$. For each dataset combination we have also reported the difference in best-fit log-likelihoods and the ratio of Bayesian evidences w.r.t. the non-interacting case NI$\nu$. Details about the models and datasets are given in section \ref{section:2}  Please note that we have opted not to present formal parameter constraints for the runs with the full range in interaction strength. The reason is that in some cases the strongly interacting region does not carry enough weight to be seen at $2\sigma$ (this is for example the case for the CMB+EXT 3$\nu$-interacting case). Whether this occurs depends on the lower boundary for the interaction strength because we use a logarithmic prior, and therefore, for the runs with the full range in interaction strength the posterior confidence regions will depend very strongly on the assumed prior. For the cases where we separate the two regions this effect is much less pronounced, which is why we choose this way of presenting our results. 
	} 

\end{table}

\begin{itemize}
	\item $\logg$ and $100\theta_{MC}$: The posterior distributions for $\logg$ and $100\theta_{MC}$ are shown in figure~\ref{fig:1} and figure~\ref{fig:2} respectively, for the runs with full prior-range of $\logg$. For both the parameters, we find the two-peak structure previously established in literature. There are, however, differences in the peak structure as we move from the 3$\nu$-interacting scenario to the 1$\nu$-interacting scenario. In the $\logg$ posteriors, the SI$\nu$ peak is centered around $\logg \simeq -1.6~ \text{to} -1.7$. The main difference between the $\logg$ posteriors is that in the 1$\nu$-interacting model the SI$\nu$ peak is far more prominent, and the posterior does not vanish between the two peaks. This happens because the cosmological data is less constraining to the 1$\nu$-interacting scenario as the interaction is limited to only one species of neutrino. From the $100\theta_{MC}$ posteriors, we see that in the 3$\nu$-interacting case the MI$\nu$ (left peaks) and SI$\nu$ (right peaks) modes are completely separated from each other (a 6.6$\sigma$ separation for the CMB+EXT dataset), whereas for the 1$\nu$-interacting case, the MI$\nu$ and SI$\nu$ peaks overlap with each other. This is again due to the fact that limiting the interaction to only one neutrino species leads to much smaller shifts in the peaks of the CMB anisotropies power spectra, which leads to overlapping MI$\nu$ and SI$\nu$ peaks.

\begin{figure}[tbp]
	\includegraphics[width=0.45\textwidth]{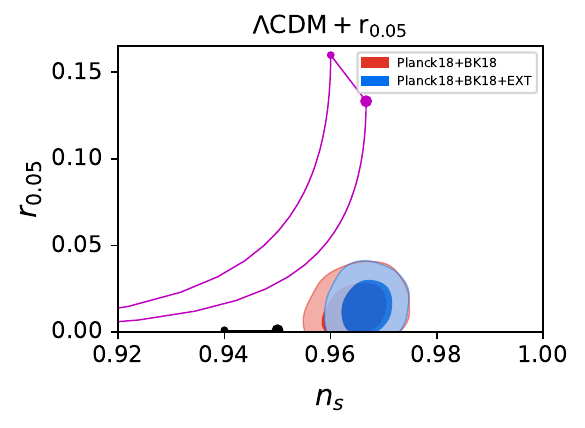}
	\hfill
	\includegraphics[width=0.45\textwidth]{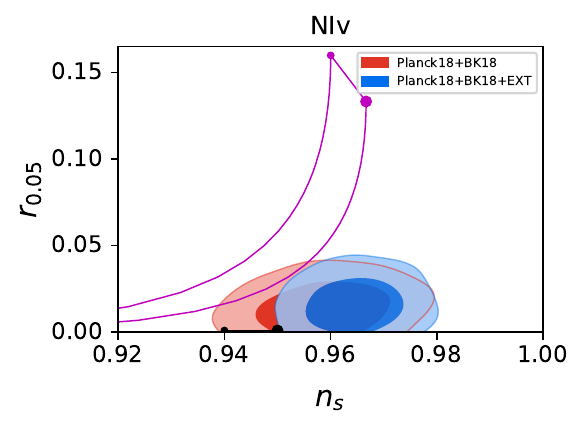}
	\includegraphics[width=0.45\textwidth]{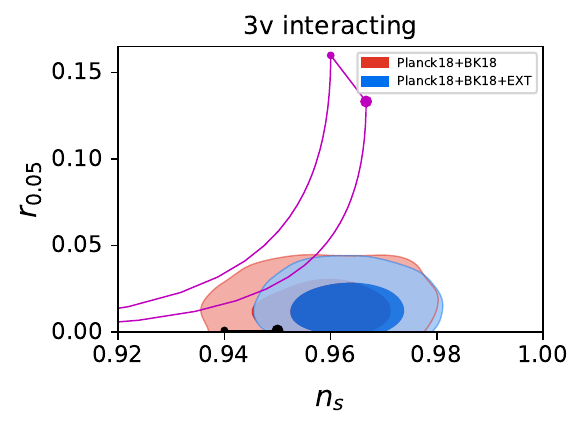}
	\hfill
	\includegraphics[width=0.45\textwidth]{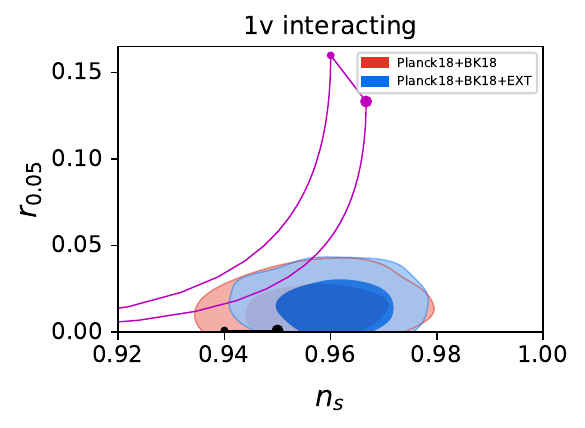}
	\caption{\label{fig:3} Here we provide 2D contour plots in the $n_s-r_{0.05}$ plane for different cosmological models. We have provided the results for two dataset combinations: Planck18+BK18 and Planck18+BK18+EXT. All the plots are from runs with full range of the coupling strength $\logg$. Apart from the 2D plots, the area covered by the magenta lines is the area predicted by Natural inflation for e-foldings $50<N_{*}<60$, whereas the black lines give the region predicted by Coleman-Weinberg inflation (which prefers a very small tensor-to-scalar ratio) for the same e-folding range. The smaller circle represents $N_{*} = 50$, whereas the larger circle represents $N_{*} = 60$. From the plots, it is clear that both in $\Lambda\textrm{CDM}+r_{0.05}$ and NI$\nu$ model, these two inflationary models are disfavoured at more that 2$\sigma$ by the Planck18+BK18+EXT data, whereas in the 3$\nu$-interacting and 1$\nu$-interacting cases, the inflationary models are allowed at 2$\sigma$, but not at 1$\sigma$. Details about models and datasets are given in section \ref{section:2}.}
\end{figure}

\begin{figure}[tbp]
	\includegraphics[width=0.45\textwidth]{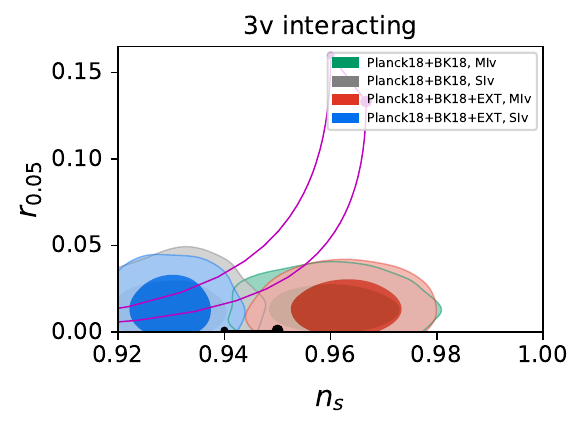}
	\hfill
	\includegraphics[width=0.45\textwidth]{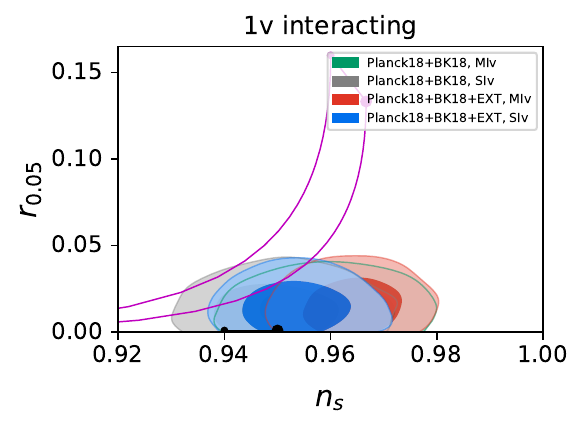}
	
	\caption{\label{fig:4} Here we provide 2D contour plots in the $n_s-r_{0.05}$ plane for the  3$\nu$-interacting and 1$\nu$-interacting cases, separately for the MI$\nu$ and SI$\nu$ modes. We have provided the results for two dataset combinations: Planck18+BK18 and Planck18+BK18+EXT. Apart from the 2D plots, the area covered by the magenta lines is the area predicted by Natural inflation for e-foldings $50<N_{*}<60$, whereas the black lines give the region predicted by Coleman-Weinberg inflation (which prefers a very small tensor-to-scalar ratio) for the same e-folding range. The smaller circle represents $N_{*} = 50$, whereas the larger circle represents $N_{*} = 60$. We see that the contours for the MI$\nu$ mode are closer to the $\Lambda\textrm{CDM} + r_{0.05}$ case, whereas the SI$\nu$ mode causes the contours to shift towards the left, with the 3$\nu$-interacting model causing a much greater shift than the 1$\nu$-interacting model. Details about models and datasets are given in section \ref{section:2}.}
\end{figure}

\item $n_s$ and $r_{0.05}$: For the runs incorporating the full prior range of $\logg$, we provide the $n_s-r_{0.05}$ 2D correlation plots in figure~\ref{fig:3} for the 3$\nu$-interacting and the 1$\nu$-interacting scenarios (the two bottom panels), along with the 
plots for the $\Lambda\textrm{CDM}+r_{0.05}$ model and the NI$\nu$ model (the two top panels). We can see that in the  $\Lambda\textrm{CDM}+r_{0.05}$ model, both the Natural Inflation (NI) and Coleman-Weingberg Inflation (CWI) models are rejected at much more than 2$\sigma$. The same is true for the PBH DM related inflationary models ($n_s \simeq 0.95$), although not shown in the figure. Incorporating the NI$\nu$ model, however, leads to an expansion of the allowed parameter space. The main reason for expansion of the allowed parameter space in the NI$\nu$ model is a strong positive correlation between $N_{\rm eff}$ and $n_s$ \cite{Gerbino:2016sgw}, as $N_{\rm eff}$ strongly affects the expansion history in the early universe. We found a correlation-coefficient of $R=+0.84$ between the two parameters in the NI$\nu$ model with the CMB+EXT dataset combination, and $R=+0.87$ with CMB-only data. There is also a small positive correlation between $\sum m_{\nu}$ and $n_s$ ($R = +0.22$ with CMB+EXT dataset), but this is not the dominant effect. But still, the NI and CWI are rejected at 2$\sigma$ when the full CMB+EXT dataset is considered. Again, the same is true for the PBH DM related inflationary models which require $n_s \simeq 0.95$ to account for all dark matter as PBHs. 

However, once we consider the 3$\nu$ and 1$\nu$-interacting models, all these inflationary models are allowed at 2$\sigma$ even with the most constraining CMB+EXT dataset. We emphasize here that lower $n_s$ values (compared to $\Lambda\rm CDM$) are preferred by the SI$\nu$ modes of the interacting models. This is clearly seen in figure~\ref{fig:4} where we provide the $n_s-r_{0.05}$ contours separately for the MI$\nu$ and SI$\nu$ models. As seen in figure~\ref{fig:4}, the 3$\nu$-interacting SI$\nu$ model causes a large shift towards left in the $n_s$ values and can comfortably accommodate the NI at 1$\sigma$, and the CWI at 2$\sigma$. On the other hand the 1$\nu$-interacting SI$\nu$ model causes a much smaller shift towards the left in $n_s$ (as the self-interaction is limited to only 1 neutrino species), but can accommodate both NI and CWI at 1$\sigma$. Thus, if future experiments find evidence for such strong interactions ($\simeq 10^9$ times stronger than weak interaction) in at least one of the neutrino species, both Natural and CW inflation can remain afloat as viable inflationary theories. At the same time, we notice that the PBH DM related inflationary models can be accommodated at 1$\sigma$ in the 1$\nu$-interacting SI$\nu$ model, but are disfavoured in the 3$\nu$-interacting SI$\nu$ model. It is also to be noted that the MI$\nu$ model in both the 3$\nu$ and 1$\nu$-interacting scenarios can accommodate all these inflationary models at 2$\sigma$.

\begin{figure}[tbp]
	\includegraphics[width=0.45\textwidth]{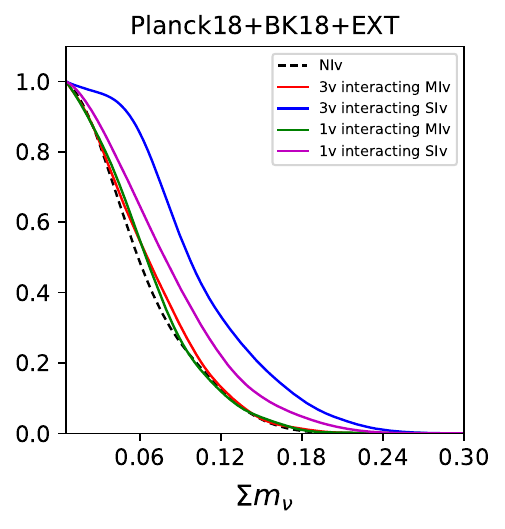}
	\hfill
	\includegraphics[width=0.45\textwidth]{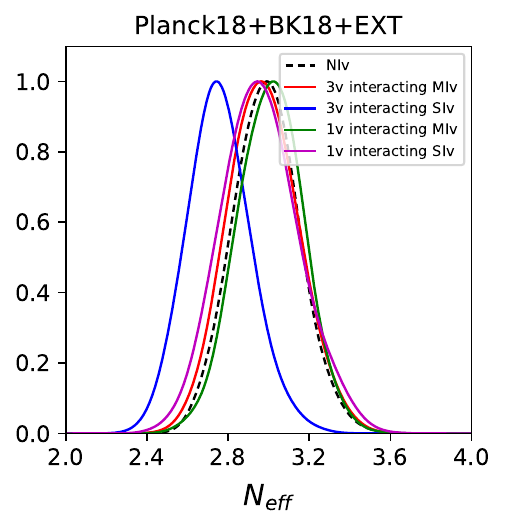}
	\caption{\label{fig:5} 1D posterior distributions of $\sum m_{\nu}$ [eV] and $N_{\rm eff}$ for the MI$\nu$ and SI$\nu$ modes separately, for the two cases: all 3 neutrinos interacting (3$\nu$-interacting), and only 1 neutrino interacting (1$\nu$-interacting). We have provided the results for the Planck18+BK18+EXT dataset combination. Details about models and datasets are given in section \ref{section:2}.}
\end{figure}

\begin{figure}[tbp]
	\includegraphics[width=0.45\textwidth]{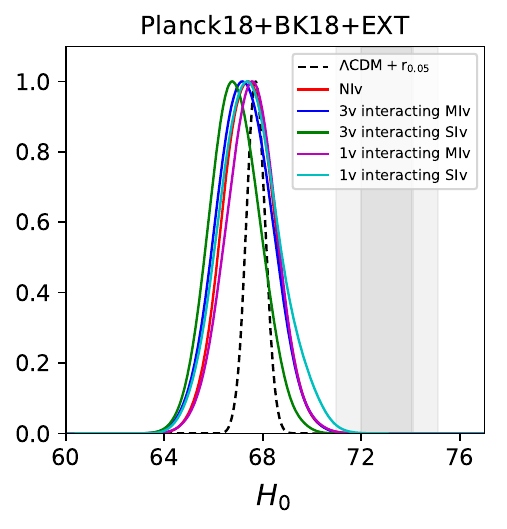}
	\hfill
	\includegraphics[width=0.45\textwidth]{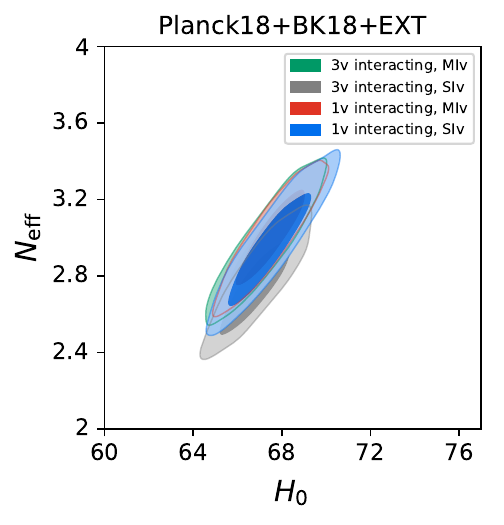}
	\caption{\label{fig:6} Left panel shows the 1D posterior distributions of $H_0$ [km/s/Mpc] for the MI$\nu$ and SI$\nu$ modes separately, for the two cases: all 3 neutrinos interacting (3$\nu$-interacting), and only 1 neutrino interacting (1$\nu$-interacting). The left panel also provides $H_0$ posteriors for the $\Lambda\textrm{CDM}+r_{0.05}$ and NI$\nu$ models. The shaded grey region corresponds to the 1$\sigma$ and 2$\sigma$ allowed regions from the local distance ladder measurement of $H_0$ from the SH0ES collaboration \cite{Riess:2021jrx}. The right panel shows the 2D correlation plot between $N_{\rm eff}$ and $H_0$. We have provided the results for Planck18+BK18+EXT dataset combination. Details about models and datasets are given in section \ref{section:2}.}
\end{figure}

\item $\sum m_{\nu}$: The 1D posterior distributions of the sum of neutrino masses parameter is given in the left panel of figure~\ref{fig:5}. We find that while the 3$\nu$-interacting and the 1$\nu$-interacting SI$\nu$ model prefers somewhat larger values of $\sum m_\nu$ than the MI$\nu$ models or the NI$\nu$ model, the obtained upper limit on $\sum m_\nu$ does not differ significantly (typically less than 20--30\%). This implies that the cosmological neutrino mass bounds quoted in literature are quite robust against the introduction of non-standard self-interactions (via a heavy mediator) in the neutrino sector. Also see \cite{RoyChoudhury:2018bsd} for the effect of CMB B-mode data on neutrino mass bounds. 

\item $N_{\rm eff}$ and $H_0$: The 1D posterior distributions of $N_{\rm eff}$ is given in the right panel of figure~\ref{fig:5}. We find that the 3$\nu$-interacting MI$\nu$ model and the 1$\nu$-interacting MI$\nu$ and SI$\nu$ model lead to similar bounds on $N_{\rm eff}$ as in the non-interacting NI$\nu$ model. However, the 3$\nu$-interacting SI$\nu$ model leads to $N_{\rm eff}$ values which are slightly lower than the NI$\nu$ model, although there is no statistically significant difference. The 1D posterior distribution of the Hubble constant ($H_0$) is given in the left panel of figure~\ref{fig:6}, whereas the 2D $N_{\textrm{eff}}-H_0$ correlation plots are given in the right panel of the same. What we find is that the inferred mean value of $H_0$ from the interacting models and the NI$\nu$ model is similar to the standard $\Lambda\textrm{CDM}+r_{0.05}$ model. However, the introduction of the new parameters leads to much larger 1$\sigma$ errors on the inferred $H_0$, which implies a reduction in the Hubble tension in the NI$\nu$ and the interacting neutrino models. From 4.9$\sigma$ in the $\Lambda\rm CDM$ model \cite{Riess:2021jrx}, the tension reduces to 3.9$\sigma$ in the NI$\nu$ model and 3$\nu$-interacting MI$\nu$ model, 4.2$\sigma$ in the 3$\nu$-interacting SI$\nu$ model, 3.8$\sigma$ in the 1$\nu$-interacting MI$\nu$ model, and 3.5$\sigma$ in the 1$\nu$-interacting SI$\nu$ model (for the CMB+EXT dataset). However, it is clear that the interacting models do not perform significantly better than the non-interacting model and they do not solve the Hubble tension. This is in agreement with previous studies that did not include tensor perturbations \cite{RoyChoudhury:2020dmd,Choudhury:2021dsc,Brinckmann:2020bcn}. 

\end{itemize}

\subsection{Model selection}
To assess the goodness of fit to the data of the MI$\nu$ and SI$\nu$ models compared to the non-interacting model NI$\nu$, we employ two different statistical methods. 

\begin{itemize}
	
	\item The Bayesian evidence ratio $Z/Z_{\text{NI}\nu}$, i.e., the Bayesian evidence for the 3$\nu$ and 1$\nu$-interacting models (MI$\nu$ and SI$\nu$ modes separately) divided by the Bayesian evidence of the non-interacting model (NI$\nu$). The calculated $Z/Z_{\text{NI}\nu}$ values are given in the table~\ref{tab:evidence}. In agreement with our previous work \cite{RoyChoudhury:2020dmd}, we see that the 3$\nu$-interacting SI$\nu$ model is disfavoured (according to Jeffrey's scale~\cite{jeffreys1998theory}) compared to the NI$\nu$ model. We had seen in \cite{RoyChoudhury:2020dmd} that the Planck 2018 high-$l$ polarization data is responsible for the disfavouring of the 3$\nu$-interacting SI$\nu$ model. For the 3$\nu$-interacting scenario, for the CMB only dataset combination, the MI$\nu$ model is slightly preferred compared to the NI$\nu$ model, whereas for the CMB+EXT dataset, it is mildly disfavoured. In the 1$\nu$-interacting scenario, both MI$\nu$ and SI$\nu$ models are only slightly preferred, for both CMB and CMB+EXT dataset combinations, to the NI$\nu$ model. Thus we see that the 1$\nu$-interacting SI$\nu$ model is not disfavoured by the cosmological data, even though the 3$\nu$-interacting SI$\nu$ model is disfavoured.

	\item The difference in log-likelihoods, i.e. $-2 \left[\log \left( \mathcal{L} / \mathcal{L}_{\text{NI}\nu} \right) \right]$, between the 3$\nu$ and 1$\nu$-interacting models (MI$\nu$ and SI$\nu$ modes separately) and the non-interacting model (NI$\nu$) at their respective best-fit points. Since the interacting models have an extra parameter, we also use the Akaike information criterion (AIC)~\cite{Akaike1975} to penalise the models for the same. It can be defined as: 
	\begin{equation}
	   \text{AIC} = 2k - 2\log\mathcal{L}\,,
	\end{equation}
	where $k$ is the number of parameters in the model. We find
	\begin{align}
	    \Delta \text{AIC} = \text{AIC}_{\text{MI}\nu} - \text{AIC}_{\text{NI}\nu} &= 2 - 2 \log\left[\mathcal{L}_{\text{MI}\nu}/\mathcal{L}_{\text{NI}\nu} \right]\,
	\end{align}  
	and similarly for SI$\nu$. Taking the values of table~\ref{tab:evidence}, we see that the 3$\nu$-interacting SI$\nu$ model leads to $\Delta \text{AIC} = 7.59$ with CMB data, and $\Delta \text{AIC} = 5.80$ with CMB+EXT, both implying that the model is disfavoured compared to the NI$\nu$ model. For all the other interacting models, $\Delta \text{AIC}$ remains positive but small, implying that the 3$\nu$-interacting MI$\nu$ model and the 1$\nu$-interacting MI$\nu$ and SI$\nu$ models are only mildly disfavoured compared to the NI$\nu$ model. 
\end{itemize}

From the two statistical considerations, we can conclude that the 3$\nu$-interacting SI$\nu$ model is disfavoured by the cosmological data, whereas the other interacting models cannot be considered so, i.e., they lead to similar fits to the data compared to the non-interacting NI$\nu$ model.

\section{Conclusions}
\label{section:4}

Inflation is theorized to give rise to the perturbations in the universe through quantum fluctuations in the inflaton field. And there are a plethora of inflationary models available in the literature. Here we consider two of them, the Natural inflation and the Coleman-Weinberg (CW) inflation, both of which are heavily disfavoured (at more than 2$\sigma$) by the current cosmological data in the standard cosmological model involving scalar and tensor perturbations, $\Lambda\textrm{CDM}+r_{0.05}$. We also consider single field inflationary models with an inflection point that can produce majority or all of dark matter as primordial black holes (PBHs). Such inflationary models require a scalar spectral index $n_s \simeq 0.95$ and are disfavoured at more than 2$\sigma$ as well. We refer to these models as PBH DM related inflationary models.

In this work, however, we consider an extension to the standard cosmological model, where we introduce self-interactions among massive neutrinos, mediated via a heavy scalar. We work in the effective 4-fermion interaction limit, and use the well established relaxation time approximation to modify the neutrino Boltzmann equations for both the scalar and tensor cases. Our baseline parameterization is $\Lambda\textrm{CDM} + r_{0.05}+\logg+ N_{\rm eff}+ \sum m_{\nu}$. We consider two different scenarios: i) all 3 neutrino spieces interacting with the same coupling strength (``3$\nu$-interacting''), ii) only 1 neutrino species interacting (``1$\nu$-interacting''). We test these scenarios against a combination of cosmological datasets. We have used the latest CMB temperature, polarisation,and lensing data from the Planck 2018 data release, the latest CMB B-mode data from the BICEP/Keck collaboration, BAO and RSD measurements from SDSS-III BOSS DR-12, additional BAO measurements from MGS and 6dFGS, and uncalibrated SNe Ia luminosity distance data from the Pantheon sample. 

The motivation towards studying this model in the context of the inflationary models is that previous studies had shown that strongly self-interacting neutrinos can accommodate a lower value of the scalar spectral index, $n_s$, than the standard cosmological model. This, in turn, means that the preferred $n_s$ and $r_{0.05}$ values in this model can allow for the aforementioned inflationary models, making them viable again. Indeed, we find, that is the case. When we vary $\logg$, in its full prior range, both the 3$\nu$-interacting and 1$\nu$-interacting models allow the Natural inflation, Coleman-Weinberg inflation, and PBH DM related inflationary models at 2$\sigma$ in the $n_s-r_{0.05}$ plane, although they are disfavoured at 1$\sigma$.  

In agreement with previous studies involving non-standard neutrino interactions with a heavy mediator, we find that the posterior of $\logg$ is bimodal and thus can be divided into two modes: a moderately interacting mode (MI$\nu$) and a strongly interacting mode (SI$\nu$). It is to be noted that the MI$\nu$ produces parameter constraints close (but not completely similar) to the non-interacting case (NI$\nu$, i.e., $G_{\rm eff} =0$), whereas it is the SI$\nu$ mode which is responsible for most of the shift in the $n_s$ parameter to lower values. When we perform our analyses by separating the modes, we find that the SI$\nu$ mode in the 3$\nu$-interacting model produces a large shift in the $n_s$, such that it can comfortably accommodate the Natural inflation within 1$\sigma$. However, we find that the 3$\nu$-interacting SI$\nu$ mode is disfavoured by the data (in terms of both Bayesian evidence and raw likelihood), whereas both modes of the 1$\nu$-interacting model produce a similar fit to the data as the non-interacting mode. The 1$\nu$-interacting SI$\nu$ mode, however, produces a smaller shift in $n_s$, which is expected considering the effects of the self-interaction are limited to one neutrino species. Interestingly, the 1$\nu$-interacting SI$\nu$ model can still accommodate both Natural and CW inflation within 1$\sigma$ for certain e-folds close to $N_{*} = 60$, and also allows the PBH DM related inflationary models at 1$\sigma$. Thus, if future experiments find evidence for such strong interactions ($\simeq 10^9$ times stronger than weak interaction) in one of the neutrino species, these inflationary models can remain afloat as viable inflationary theories. Meanwhile, both the 3$\nu$-interacting and 1$\nu$-interacting MI$\nu$ models can accommodate all these inflationary models at 2$\sigma$.

Lastly, we note that the NI$\nu$, 3$\nu$-interacting, and the 1$\nu$-interacting models can partially reduce the Hubble tension, although they do not solve it. Moreover, the interacting models do not perform significantly better than the non-interacting model, in agreement with recent studies that used only scalar perturbation equations \cite{RoyChoudhury:2020dmd,Choudhury:2021dsc,Brinckmann:2020bcn}.

\section*{Acknowledgements}
We thank Arnab Paul and Guillermo Ballesteros for useful discussions regarding inflationary models. We acknowledge the use of the HPC facility at Aarhus University (http://www.cscaa.dk/) for all the numerical analyses done in this work. TT was supported by a research grant (29337) from VILLUM FONDEN.

\appendix

\bibliographystyle{utcaps}
\bibliography{references}

\end{document}